\documentclass[12pt]{spieman}  
\usepackage{amsmath,amsfonts,amssymb}
\usepackage{graphicx}
\usepackage{setspace}
\usepackage{tocloft}
\usepackage{siunitx}
\usepackage{comment}
\usepackage[left]{lineno}
\title{X-ray Radiation Damage Effects on Double-SOI Pixel Detectors for the Future Astronomical Satellite {``FORCE''}}

\author[a]{Masatoshi~Kitajima}
\author[b,a,*]{Kouichi~Hagino}
\author[a]{Takayoshi~Kohmura}
\author[a]{Mitsuki~Hayashida}
\author[a]{Kenji~Oono}
\author[a]{Kousuke~Negishi}
\author[a]{Keigo~Yarita}
\author[a]{Toshiki~Doi}
\author[a]{Shun~Tsunomachi}
\author[c]{Takeshi~G.~Tsuru}
\author[c]{Hiroyuki~Uchida}
\author[c]{Kazuho~Kayama}
\author[c]{Ryota~Kodama}
\author[d]{Takaaki~Tanaka}
\author[e]{Koji~Mori}
\author[e]{Ayaki~Takeda}
\author[e]{Yusuke~Nishioka}
\author[e]{Masataka~Yukumoto}
\author[e]{Kira~Mieda}
\author[e]{Syuto~Yonemura}
\author[e]{Tatsunori~Ishida}
\author[f]{Yasuo~Arai}
\author[g]{Ikuo~Kurachi}
\affil[a]{Tokyo University of Science, School of Science and Technology, Department of Physics, 2641 Yamazaki, Noda, Chiba, Japan, 278-8510}
\affil[b]{Kanto Gakuin University, Research Advancement and Management Organization, 1-50-1 Mutsuura-higashi, Kanazawa-ku, Yokohama, Japan, 236-8501}
\affil[c]{Kyoto University, Faculty of Science, Department of Physics, Kitashirakawa-Oiwakecho, Sakyo-ku, Kyoto, Japan, 606-8502}
\affil[d]{Konan University, Department of Physics, 8-9-1 Okamoto, Higashinada, Kobe, Hyogo, Japan, 658-8501}
\affil[e]{University of Miyazaki, Faculty of Engineering, Department of Applied Physics, 1-1 Gakuen-Kibanodai-Nishi, Miyazaki, Miyazaki, Japan, 889-2192}
\affil[f]{High Energy Accelerator Research Organization (KEK), Institute of Particle and Nuclear Studies, 1-1 Oho, Tsukuba, Ibaraki, Japan, 305-0801}
\affil[g]{D\&S Inc., 774-3-213 Higashiasakawacho, Hachioji, Tokyo, Japan, 193-0834}

\cftpagenumbersoff{figure}
\cftpagenumbersoff{table} 
\begin{document} 
\maketitle

\begin{abstract}
We have been developing the monolithic active pixel detector ``XRPIX'' onboard the future X-ray astronomical satellite ``FORCE''. XRPIX is composed of CMOS pixel circuits, SiO$_{2}$ insulator, and Si sensor by utilizing the silicon-on-insulator (SOI) technology. When the semiconductor detector is operated in orbit, it suffers from radiation damage due to X-rays emitted from the celestial objects as well as cosmic rays. From previous studies, positive charges trapped in the SiO$_{2}$ insulator are known to cause the degradation of the detector performance. To improve the radiation hardness, we developed XRPIX equipped with Double-SOI (D-SOI) structure, introducing an additional silicon layer in the SiO$_{2}$ insulator. This structure is aimed at compensating {for} the effect of the trapped positive charges. Although the radiation hardness to cosmic rays of the D-SOI detectors has been evaluated, {the} radiation effect due to the X-ray irradiation has not been evaluated. Then, we conduct an X-ray irradiation experiment using an X-ray generator with a total dose of 10~krad at the SiO$_{2}$ insulator, equivalent to 7 years in orbit. As a result of this experiment, the energy resolution in full-width half maximum for the 5.9~keV X-ray degrades by ${\rm17.8 \pm 2.8\%}$ and the dark current increases by ${\rm 89\pm 13\%}$. We also investigate the physical mechanism of the increase in the dark current {due to X-ray irradiation} using TCAD simulation. It is found that the increase in the dark current can be explained by the increase in the interface state density at the Si/SiO$_{2}$ interface.

\end{abstract}

\keywords{X-ray astronomy; silicon-on-insulator; X-ray detectors; radiation damage; dark current; surface recombination}


{\noindent \footnotesize\textbf{*}Address all correspondence to Kouichi Hagino,  \linkable{hagino@kanto-gakuin.ac.jp} }

\begin{spacing}{1}   

\section{Introduction}
\label{sect:intro}  
Since the 1990s, the charge-coupled device (CCD) has been the standard detector used in X-ray astronomy satellites. X-ray CCD has {an} excellent position and energy resolution, however, it has {a} poor time resolution of a few seconds and a narrow observable energy band of $0.3\textrm{--}10$ keV. In order to realize broadband and high-sensitivity X-ray observation, we have been developing the monolithic active pixel detector ``XRPIX'' onboard the future X-ray astronomical satellite ``FORCE''~\cite{mori2016,nakazawa2018}. FORCE will be equipped with two X-ray super-mirrors, {whose} angular resolution will be better than $15''$ in {half-power diameter}. The focal plane detector is composed of {two} stacks of Si sensors (XRPIX) and CdTe sensors and these detectors cover the X-ray energy ranging from {1~keV to 79~keV}. XRPIX is composed of CMOS pixel circuits,  SiO$_{2}$ insulator called buried oxide (BOX) layer, and Si sensor by utilizing the silicon-on-insulator (SOI) technology~\cite{tsuru2018}. This makes it possible to implement Si with low and high resistivity in the circuit and sensor layers, respectively. Since the sensor layer has high resistivity, the depletion layer thickness can be a few hundreds of $\mathrm{\mu m}$. In addition, the CMOS circuit in each pixel has a self-trigger function, so that only the pixels where an X-ray is incident can be read out, achieving a time resolution of $<10~\mathrm{\mu s}$.

XRPIX will suffer from the radiation damage due to the irradiation of bright {X-rays} from compact stars as well as {high-energy} cosmic rays. When the detector is irradiated by charged particles or X-rays, electron-hole pairs are produced in Si and also in SiO$_2$. In the former case, carriers are collected by electrodes. On the other hand, in SiO$_2$ insulator, electrons are immediately collected to electrodes, but some {parts} of holes are trapped there because of their low mobility $\mu_{h}$ compared with that of electrons $\mu_{e}$ ($\mu_{h}/\mu_{e}\sim 10^{-10}$ at $300~\mathrm{K}$)~\cite{hughes2008}. It is known that the trapped positive charges in SiO$_2$ insulator cause a shift of threshold voltages of CMOS pixel circuits and degradation of the detector performance~\cite{hara2019}.

\label{sect:expt}
\hypertarget{6C_crosssection}{}
\begin{figure}[tbp]
\begin{center}
\begin{tabular}{c}
\includegraphics[width=0.8\hsize]{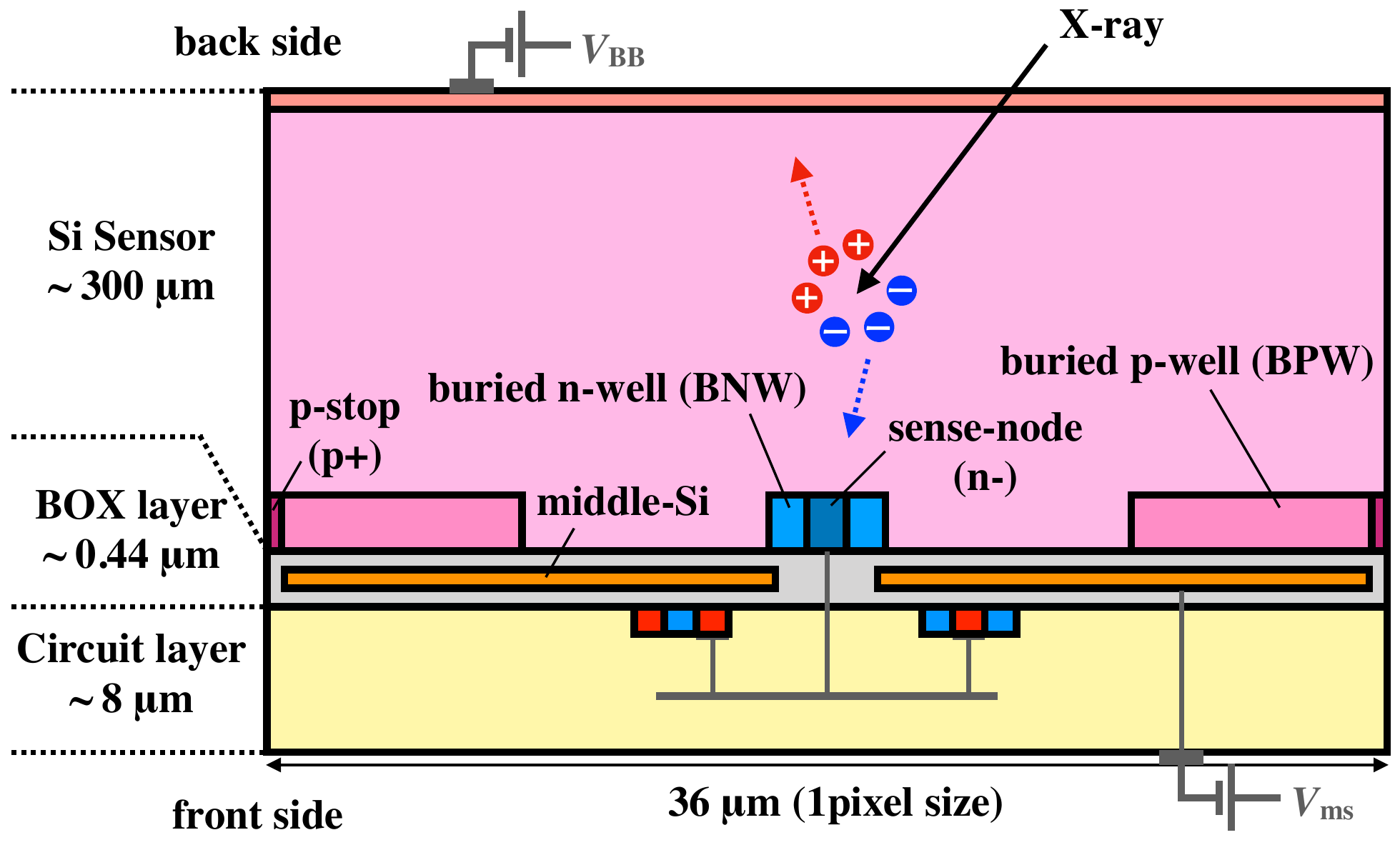}
\end{tabular}
\end{center}
\caption{Schematic cross sectional view of XRPIX6C with D-SOI structure.}
\label{6C_crosssection}
\end{figure}

In order to improve the radiation hardness, we introduced a Double SOI (D-SOI) structure, which has an additional Si layer called ``middle-Si'' in the SiO$_2$ insulator as shown in Fig.~\hyperlink{6C_crosssection}{\ref{6C_crosssection}}. It is effective against the radiation damage because the negative voltage applied on the middle-Si negates the effect of trapped positive charges in SiO$_2$ insulator caused by the radiation exposure~\cite{hagino2020}.

In the past experiment, in order to evaluate the effect of cosmic rays mainly composed of high energy protons, we evaluated the radiation hardness of D-SOI detector by 6 MeV proton beam irradiation~\cite{hagino2020}.
{In this experiment, we found that even after irradiation of $\sim 5{\rm ~krad}$, degradation of the energy resolution was as small as 7\%. Moreover, we found that the gain degradation can be quantitatively explained by the sense-node capacitance increased by the trapped positive charges.
On the other hand, since high-energy X-rays up to $79~\mathrm{keV}$ emitted from celestial objects are focused with X-ray super-mirrors at the focal point with a high angular resolution of $<15''$, the focused high-energy X-rays can cause serious radiation damage against XRPIX~\cite{mori2019}.}
Thus, {in this paper}, we conducted an X-ray irradiation experiment {to evaluate} the X-ray radiation hardness of D-SOI detector~\cite{kitajima2021}.
The X-ray irradiation experiment is described in Sec.~\ref{sect:expt} and we show its results in Sec.~\ref{sect:result}. In Sec.~\ref{sect:disc}, we discuss the possible cause of the degradation of detector performance mainly about the dark current using device simulation. Section~\ref{sect:conc} provides the conclusions of this study. 

\section{X-ray Irradiation Experiment}
We conducted an X-ray irradiation experiment on D-SOI detector, called ``XRPIX6C''. Figure~\hyperlink{6C_crosssection}{\ref{6C_crosssection}} and Table~\hyperlink{chip_design_of_6C}{\ref{chip_design_of_6C}} show the schematic cross-sectional structure and chip design of XRPIX6C, respectively. The sensor layer, whose thickness is $300~\mathrm{\mu m}$, is p-type Si bulk and its resistivity is $4~\mathrm{k\si{\ohm}~cm}$. With this thickness, the {back-bias voltage} $V_\mathrm{BB}$ should be higher than $\simeq -216~\mathrm{V}$ for full depletion. In this experiment, we applied a {back-bias voltage} $V_\mathrm{BB}=-250~\mathrm{V}$. In this device, each pixel is isolated from each other by a p-stop and has a sense node surrounded by a buried n-well (BNW). BNW was introduced to prevent interference between {the} sensor layer and circuit layer~\cite{arai2011}. Buried p-well (BPW) was introduced to generate a lateral electric field structure from {the} pixel boundary to the sense node so that electric charges collect at the sense node. It is also effective in suppressing dark current by covering the Si/SiO$_{2}$ interface. It is the reason why we call this device ``Double SOI'' that {an} additional Si layer called middle-Si is introduced in the BOX layer. It compensates for the effect of positive charges trapped in the BOX layer by biasing negatively. We applied a negative voltage of $-2.5~\mathrm{V}$ to middle-Si during the experiment.

\hypertarget{chip_design_of_6C}{}
\begin{table}[tbp]
\caption{The chip design of XRPIX6C.} 
\label{chip_design_of_6C}
\begin{center}
\begin{tabular}{|l|l|} 
\hline
\rule[-1ex]{0pt}{3.5ex}  parameter & value  \\
\hline\hline
\rule[-1ex]{0pt}{3.5ex}  Chip size & $4.45~\mathrm{mm}\times 4.45~\mathrm{mm}$  \\
\hline
\rule[-1ex]{0pt}{3.5ex}  Sensor area & $1.7~\mathrm{mm}\times 1.7~\mathrm{mm}$    \\
\hline
\rule[-1ex]{0pt}{3.5ex}  Pixel size & $36~\mathrm{\mu m}\times 36~\mathrm{\mu m}$    \\
\hline
\rule[-1ex]{0pt}{3.5ex}  Number of pixels & $48\times 48$    \\
\hline
\rule[-1ex]{0pt}{3.5ex}  Thickness of sensor & 300 $\mathrm{\mu m}$  \\
\hline
\rule[-1ex]{0pt}{3.5ex}  Type of sensor layer & Froating zone p-type Si  \\
\hline
\rule[-1ex]{0pt}{3.5ex}  Sensor resistivity & $4~\mathrm{k\si{\ohm}~cm}$  \\
\hline
\end{tabular}
\end{center}
\end{table} 

\hypertarget{xraydamage_setup}{}
\begin{figure}[tbp]
\begin{center}
\begin{tabular}{c}
\includegraphics[width=0.8\hsize]{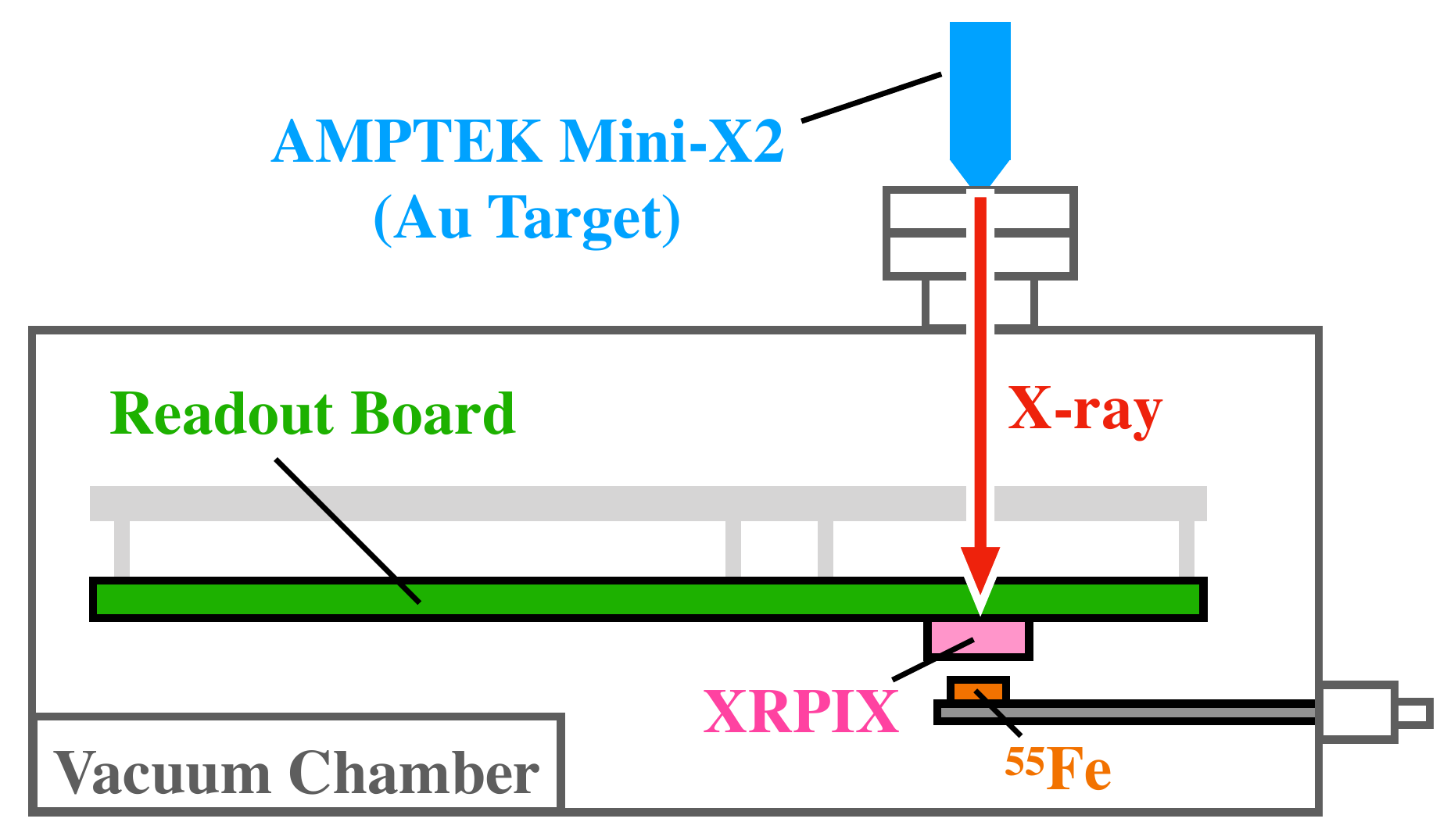}
\end{tabular}
\end{center}
\caption[Schematic view of experimental set up.]{Schematic view of experimental set up. We irradiated X-rays on the back side of XRPIX by using an X-ray tube, and irradiated X-rays of $^{55}$Fe on the front side to evaluate the performance of XRPIX.}
\label{xraydamage_setup}
\end{figure}

Figure~\hyperlink{xraydamage_setup}{\ref{xraydamage_setup}} shows the schematic view of our experimental setup. XRPIX6C was installed in a vacuum chamber and cooled down to $\simeq -65 {}^\circ \mathrm{C}$ in order to reduce the shot noise of the dark current. We irradiated X-rays on the back side (sensor layer side; see Fig.~\hyperlink{6C_crosssection}{\ref{6C_crosssection}}) of XRPIX6C by using an X-ray tube (Mini-X2, AMPTEK) attached to the vacuum chamber. The X-ray tube was operated at $20~\mathrm{kV}$ {with a target of Au}. {The energies of L-shell fluorescence lines of Au are 9.7~keV (L$_\alpha$), 11.4~keV (L$_\beta$), and 13.4~keV (L$_\gamma$).} XRPIX6C was irradiated with X-ray to {a} total dose of $10~\mathrm{krad}$ at the BOX layer. Assuming that we observe the Crab Nebula, one of the observational targets of FORCE, for $100~\mathrm{ksec}$ per month based on a  previous study~\cite{mori2019}, $10~\mathrm{krad}$ corresponds to $7$ years in-orbit operation.

The degradation of detector performance {was} monitored by iterating the X-ray irradiation and data acquisition of $^{55}$Fe. 
{These evaluation data were taken after irradiations of 0.1, 0.5, 1, 2, 4, 7, and 10~krad. In order to avoid the large dark current after the irradiation, the evaluations were performed after the dark current settled down to steady values.}

\section{Results of Irradiation Experiment}
\label{sect:result}

\hypertarget{spec_ADU}{}
\begin{figure}[tbp]
\begin{center}
\begin{tabular}{c}
\includegraphics[width=0.8\hsize]{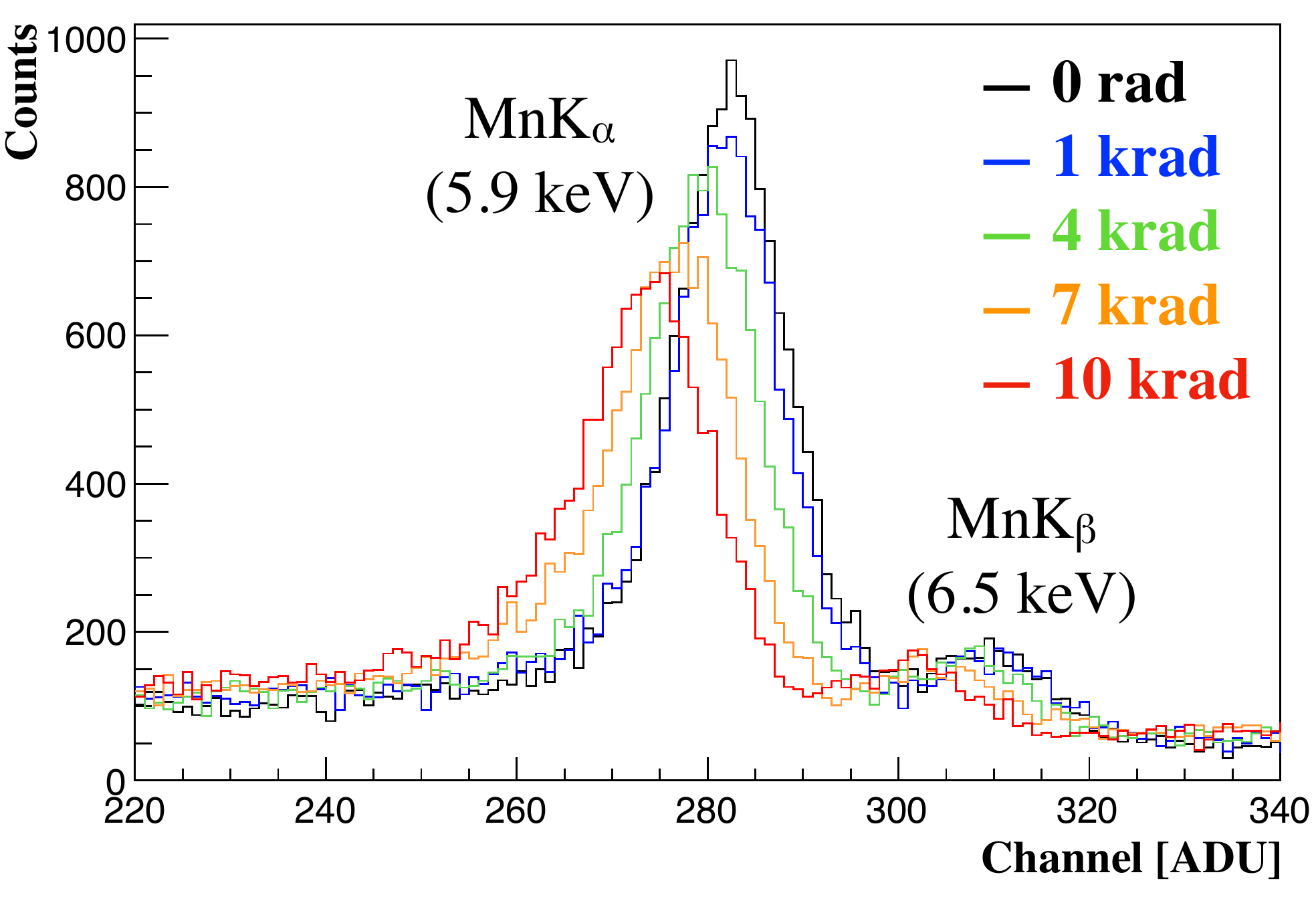}
\end{tabular}
\end{center}
\caption[${}^{55}$Fe energy spectra of XRPIX6C before and after X-ray irradiation.]{${}^{55}$Fe energy spectra of XRPIX6C before and after X-ray irradiation. The horizontal axis is uncorrected pulse height in Analog-to-Digital Unit (ADU) and the vertical axis is the number of counts.}
\label{spec_ADU}
\end{figure} 

We evaluated the spectral performance of XRPIX by irradiating X-rays of the ${}^{55}\mathrm{Fe}$ radioisotope from the front side (circuit layer side). {We took $5\times10^5$ frames of the X-ray data with 1-ms integration time.} Figure~\hyperlink{spec_ADU}{\ref{spec_ADU}} shows the spectra of {single-pixel events, which are} extracted when a pulse height of one pixel exceeds the event threshold {($100{\rm ~ADU}\simeq2.1{\rm~keV}$)}, while pulse heights of the surrounding 8 pixels are below the split threshold {($60{\rm ~ADU}\simeq1.3{\rm~keV}$)}. The horizontal axis is uncorrected pulse {heights} in Analog-to-Digital Unit (ADU) and the vertical axis is the number of counts. The tail structure {on the low-energy side of the peak} becomes noticeable as the dose increases. This is probably caused by the charge loss due to the increase in the interface trap at the Si/SiO$_{2}$ interface by X-ray irradiation~\cite{hagino2019}.

\begin{figure}[tbp]
\begin{center}
\begin{tabular}{c}
\includegraphics[width=\hsize]{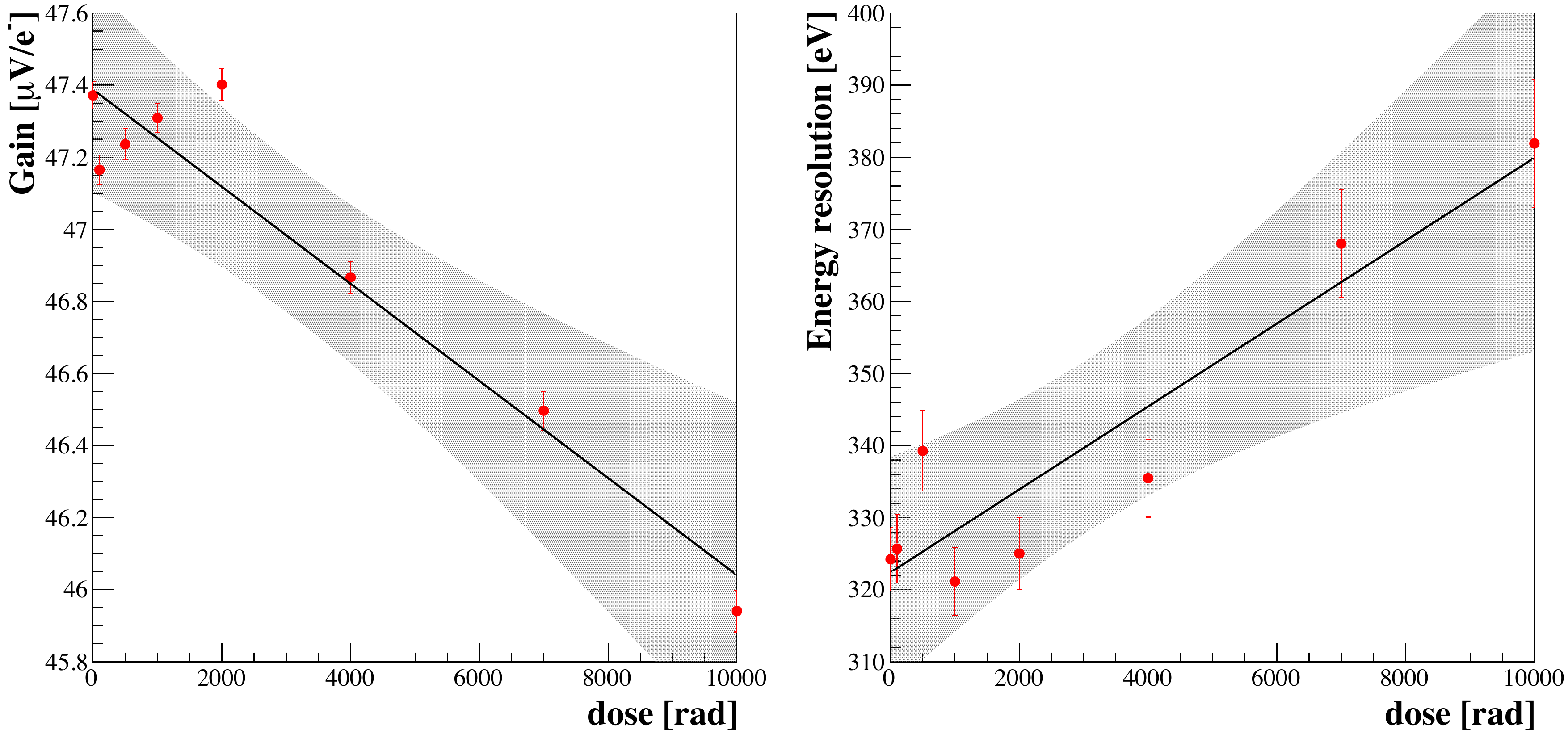}
\\
(a) \hspace{7.5cm} (b)
\end{tabular}
\end{center}
\caption[Conversion gain and energy resolution of XRPIX6C.]{(a) Conversion gain and (b) energy resolution of XRPIX6C as a function of dose level. Solid lines and shaded regions indicate the best fit linear functions and 95\% confidence intervals, respectively. {The confidence intervals are calculated on the assumption of Gaussian uncertainty of the data.}}
\label{Gain_FWHM}
\end{figure}

We also evaluated the conversion gain and energy resolution of XRPIX6C by using the measured peak position and FWHM of the MnK$_{\alpha}$ line derived by fitting with {the} Gaussian function. Figure~\hyperlink{Gain_FWHM}{\ref{Gain_FWHM}(a)} and~\hyperlink{Gain_FWHM}{\ref{Gain_FWHM}(b)} show the gain and energy resolution as {functions} of the total dose, respectively. Both of them were almost constant up to $2~\mathrm{krad}$, but after that, they degraded with increasing dose. {Although an outlier of the gain at 2~krad indicates a possibility of a non-linear relation to the dose, we assume a linear relation for simplicity.} After $10~\mathrm{krad}$ irradiation, the gain and energy resolution degraded by $2.84\pm 0.34\%$ and $17.8\pm 2.8\%$ compared with that of non-irradiation, respectively.

\hypertarget{Gain_FWHM}{}

\begin{figure}[tbp]
\begin{center}
\begin{tabular}{c}
\includegraphics[width=\hsize]{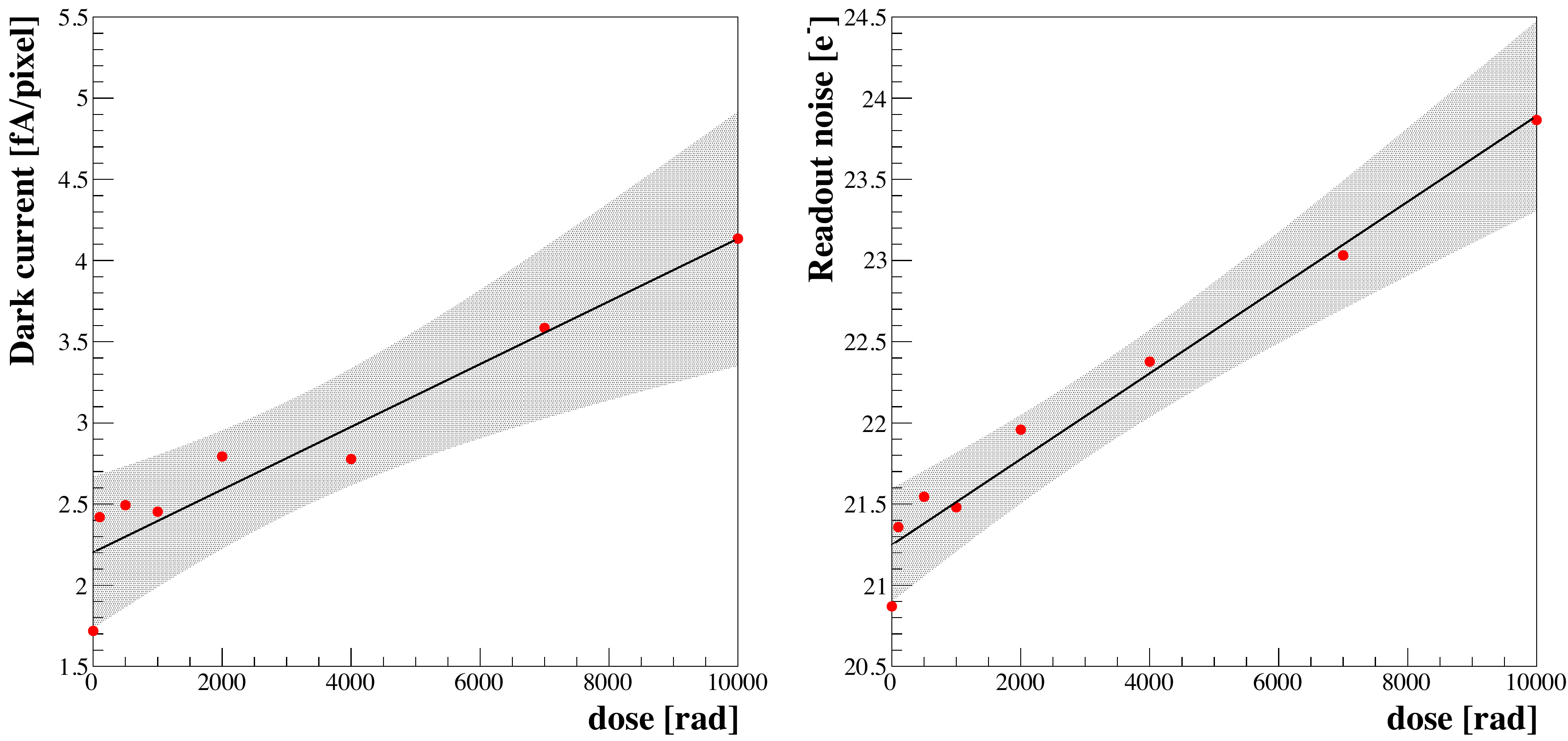}
\\
(a) \hspace{7.5cm} (b)
\end{tabular}
\end{center}
  \caption[Dark current and readout noise of XRPIX6C  as a function of dose level.]{(a) Dark current and (b) readout noise of XRPIX6C as a function of dose level. Best fit linear functions and 95\% confidence intervals are overplotted similarly as in Fig.~\hyperlink{Gain_FWHM}{\ref{Gain_FWHM}}.}
\label{Leak_Nro}
\end{figure}

Figure~\hyperlink{Leak_Nro}{\ref{Leak_Nro}(a)} and~\hyperlink{Leak_Nro}{\ref{Leak_Nro}(b)} show the dark current and readout noise as {functions} of the dose, respectively. {In order to evaluate the dark current, we measured the pedestal levels as a function of the integration time by reading pulse heights from all the pixels with integration times of 0.1, 1, 2, 4, 8, 16, and 32~ms. Although this measurement was performed under the irradiation of X-rays from $^{55}$Fe, its count rate was much less than 0.1~count/frame/pixel, even with the longest integration time of 32~ms. Thus, the effect of X-ray events on the pedestal measurement was negligible. In this measurement,} the longer the integration time, the more charge is accumulated and the higher pedestal level is output. {Thus,} we estimated the relationship between the integration time and pedestal level as a linear function, and evaluated the dark current from the slope of the function. In addition, we evaluated the readout noise by measuring {the} pedestal width of each pixel. As shown in Fig.~\hyperlink{Leak_Nro}{\ref{Leak_Nro}}, both of them increased in proportion to the dose. After $10~\mathrm{krad}$ irradiation, the dark current and readout noise increased by $89\pm 13\%$ and $12.4\pm 0.9\%$ compared with that of non-irradiation, respectively.

\hypertarget{Leak_Nro}{}

\begin{figure}[tbp]
\hypertarget{Noisypix}{}
\begin{minipage}[t]{0.48\hsize}
\begin{center}
\includegraphics[width=\hsize]{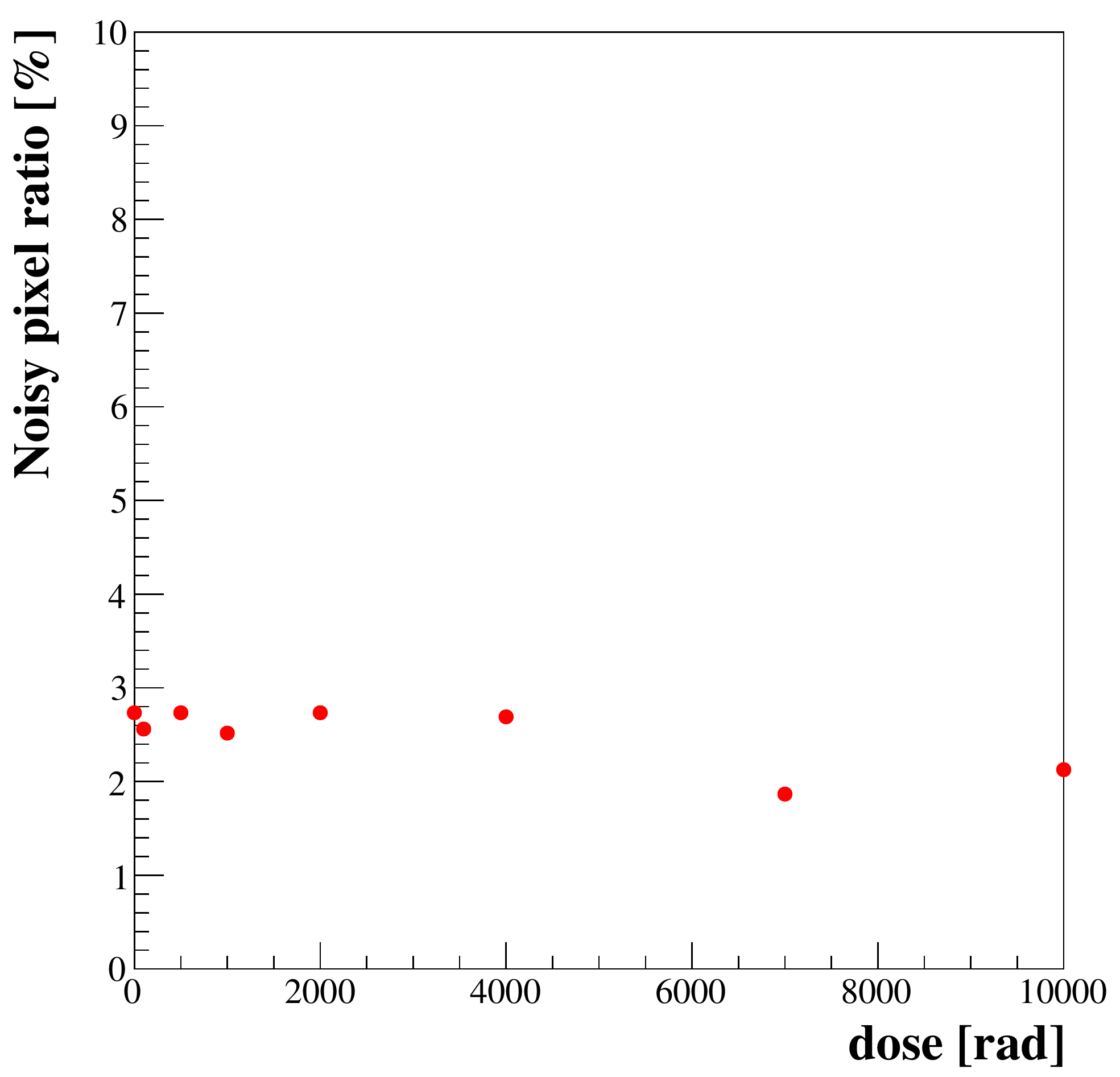}
\end{center}
\caption[Noisy pixel ratio  of XRPIX6C as a function of dose level.]{Noisy pixel ratio  of XRPIX6C as a function of dose level.}
\label{Noisypix}
\end{minipage}
\hspace{0.01\hsize}
\hypertarget{Noisypix_comp}{}
\begin{minipage}[t]{0.48\hsize}
\begin{center}
\includegraphics[width=\hsize]{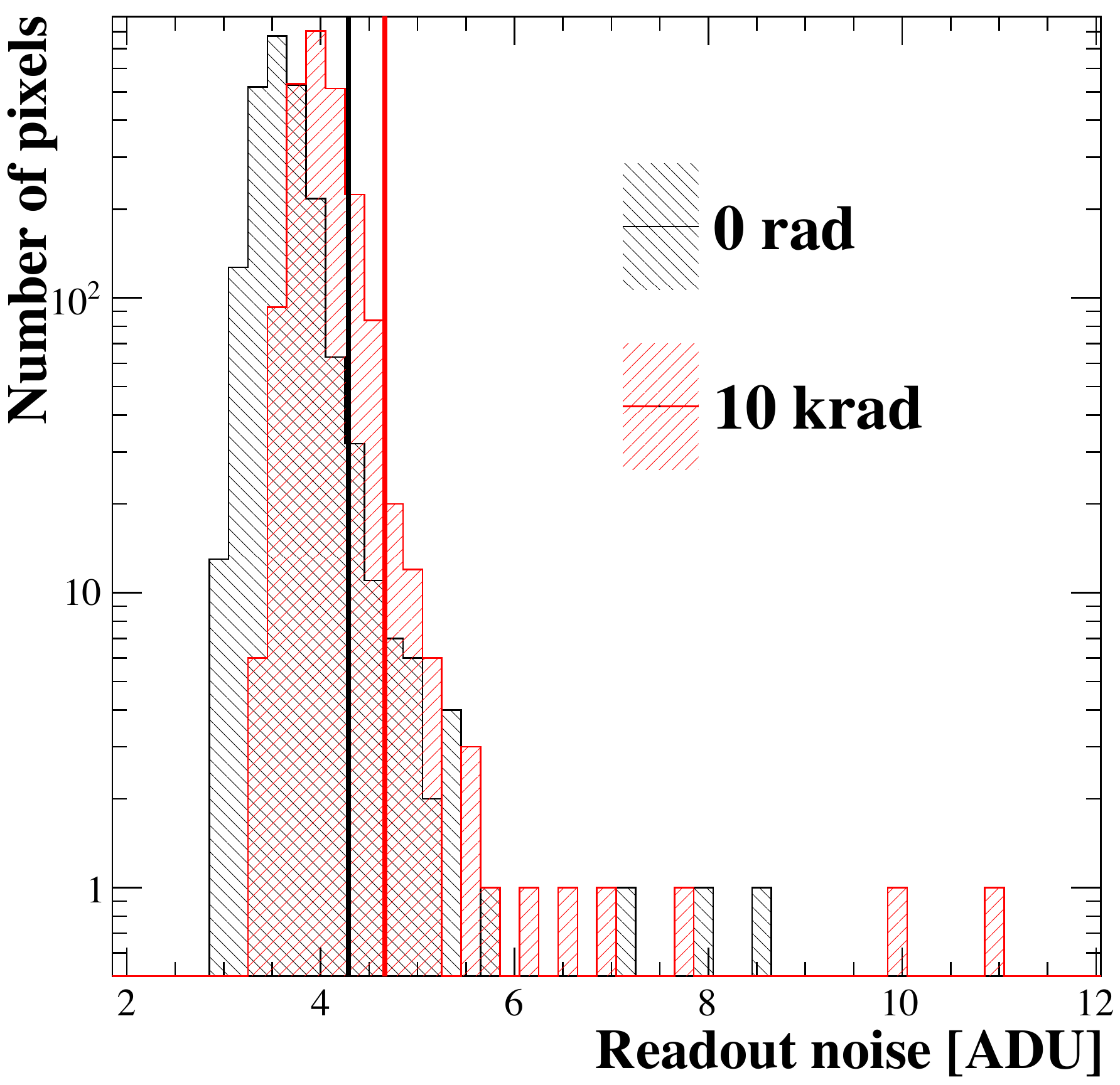}
\end{center}
\caption[Distributions of readout noise in each pixel of XRPIX6C before and after the X-ray irradiation.]{Distributions of readout noise in each pixel of XRPIX6C before and after the X-ray irradiation. The noisy-pixel thresholds at 3-$\sigma$ are shown in black (0~rad) and red (10~krad) vertical thick solid lines.}
\label{Noisypix_comp}
\end{minipage}
\end{figure}

We also evaluated the fraction of ``noisy pixels'' to all pixels. The pixels above $3\sigma$ in the histogram of readout noise is judged to be noisy pixels. As shown in Fig.~\hyperlink{Noisypix}{\ref{Noisypix}}, there was little change from non-irradiation to $10~\mathrm{krad}$. {Since the noisy pixels are not remarkably different from the normal pixels, the noisy pixels at 0~krad were not exactly the same pixels as those at 10~krad. As shown in Fig.~\hyperlink{Noisypix_comp}{\ref{Noisypix_comp}}, distributions of the readout noises in each pixel were almost the same shape with a slight shift. Therefore, these results} suggest that there is an increase in noise on average, but no pixels show any extreme increase in noise. 

\section{Discussion}
\label{sect:disc}
\subsection{Energy resolution}
\label{subsect:disc_fwhm}

\hypertarget{spec_keV}{}
\begin{figure}[tbp]
\begin{center}
\begin{tabular}{c}
\includegraphics[width=0.8\hsize]{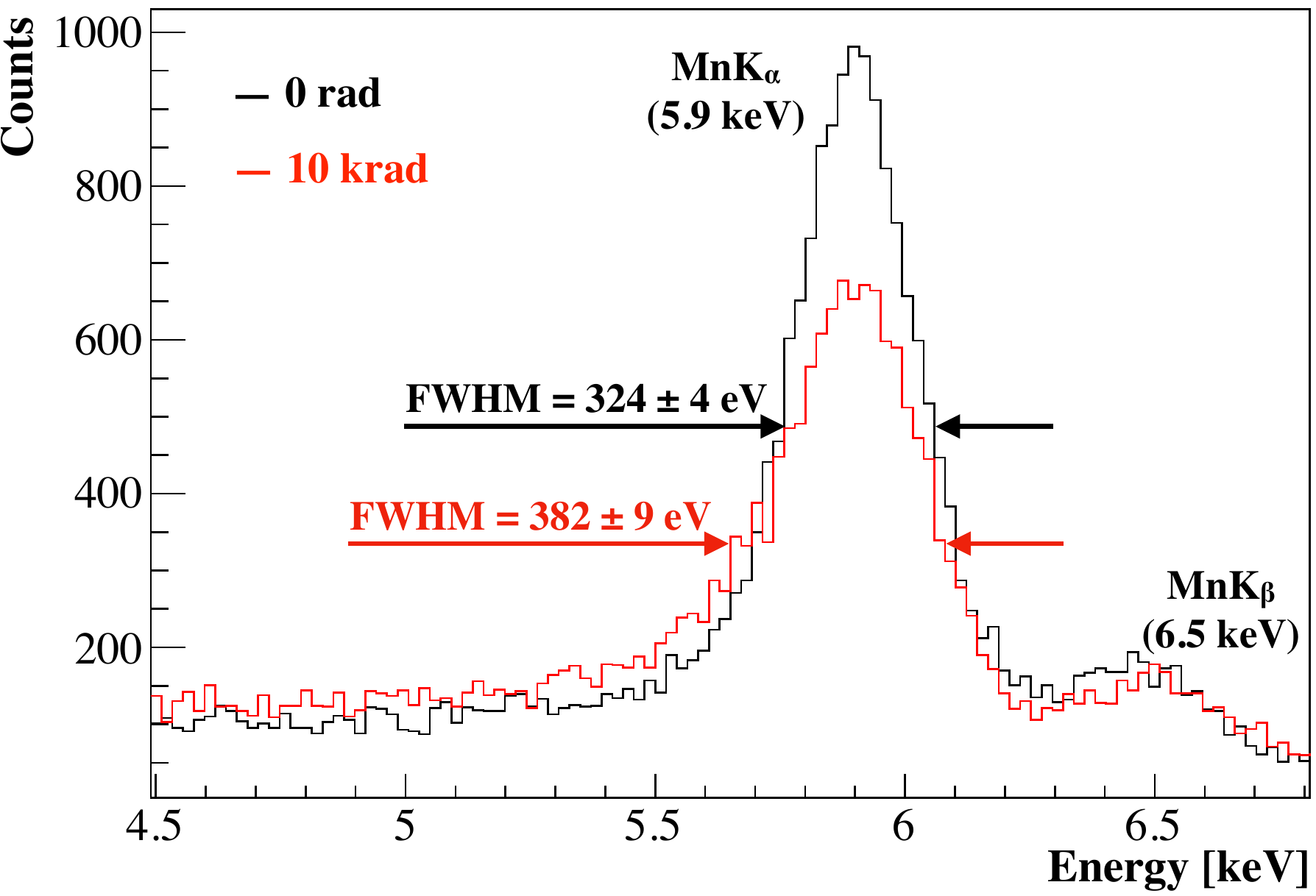}
\end{tabular}
\end{center}
\caption[${}^{55}$Fe energy spectra corrected for the gain degradation before and after X-ray irradiation.]{${}^{55}$Fe energy spectra of XRPIX6C corrected for the gain degradation before and after X-ray irradiation. The horizontal axis is the X-ray energy and the vertical axis is the number of counts. The tail structure increased at $10~\mathrm{krad}$ compared with pre-irrad.}
 \label{spec_keV}
\end{figure} 

Figure~\hyperlink{spec_keV}{\ref{spec_keV}} shows the ${}^{55}$Fe energy spectra corrected for the gain degradation at $0~\mathrm{rad}$ and $10~\mathrm{krad}$. The horizontal axis is the X-ray energy and the vertical axis is the number of counts. In Fig.~\hyperlink{spec_keV}{\ref{spec_keV}}, the peak positions are aligned in order to focus on the change in spectral shape rather than the gain degradation. The increase in the tail structure of the X-ray spectra was obviously observed after $10~\mathrm{krad}$ irradiation. It contributes {to} the degradation of spectral performance.

This tail structure can be seen even before the irradiation, and we investigated this issue in the previous study~\cite{hagino2019}. This effect is likely caused by the charge loss at the Si/SiO$_{2}$ interface. When carriers generated in the sensor layer drift towards the sense-node along the electric field, they pass through the Si/SiO$_{2}$ interface, and some of them are captured in the trap level. {As anticipated from prior studies, the interface state density increases due to radiation damage\cite{Schwank2008}, increasing the effects of charge loss and leading to an enlargement of the tail structure observed in Fig.~\hyperlink{spec_keV}{\ref{spec_keV}}.}

\subsection{Gain}
\label{subsect:disc_gain}
The chip output gain degraded by $2.84\pm 0.34$\% after $10~\mathrm{krad}$ irradiation. According to the previous study, the gain degradation due to the radiation damage is caused by the enlargement of BNW~\cite{hagino2020}. This is explained by the effect of the positive charge trapped in the BOX layer due to X-ray irradiation. Its potential attracts electrons and enlarges the area of BNW. This phenomenon results in an increase in the sense node capacitance and the degradation of the gain. The relation between the inverse of the gain $G$ and BNW size $S_\mathrm{BNW}$ is described in the previous study~\cite{hagino2020} {as
\begin{equation}
\Delta \left(\frac{1}{G}\right)\simeq 3.4\times 10^{-3}\times \left(\frac{\Delta S_\mathrm{BNW}}{1~\mathrm{\mu m}^2}\right)~\mathrm{fF}.
 \label{gain_bnw}
\end{equation}
}
According to this equation, the change in the inverse of the gain $\Delta \left(1/G\right)\simeq 0.11~\mathrm{fF}$ after $10~\mathrm{krad}$ irradiation is equivalent to the enlargement of BNW by $\Delta S_\mathrm{BNW}\simeq 31~\mathrm{\mu m}^2$. As the BNW width $w_\mathrm{BNW}$ is designed to be $3~\mathrm{\mu m}$, $w_\mathrm{BNW}\simeq\sqrt{3^2+31}\simeq 6.3~\mathrm{\mu m}$ at $10~\mathrm{krad}$. It is {a} reasonable value because the distance between BNW and BPW is designed to be $7~\mathrm{\mu m}$. In addition, it is considered that the charge loss at the Si/SiO$_{2}$ interface due to the increase in the interface state density contributed to the peak shift and caused the gain degradation.

\subsection{Readout noise}
\label{subsect:disc_ron}
The readout noise increased by $12.4\pm 0.9\%$ after $10~\mathrm{krad}$ irradiation. The gain degradation discussed in the previous section affects the increase of the readout noise. {According to a previous study~\cite{harada2019},} the readout noise $\sigma$ in XRPIX is related to the gain $G$ with {an empirical relation} of $\sigma \propto G^{-0.7}$. Therefore, the gain degradation of $\simeq 2.8\% $ after $10~\mathrm{krad}$ contributes to the increase in the readout noise by $\simeq 2.2\% $. In addition, the increase in the shot noise due to the dark current increase also contributes to the increase in the readout noise. As the readout noise and dark current were evaluated using an integration time of $1~\mathrm{ms}$, the dark current increase of $\simeq 89\% $ after $10~\mathrm{krad}$ contributes to the increase in the readout noise by $\simeq 1.7\% $. Therefore, it is assumed that the gain degradation and shot noise increase due to increased dark current do not contribute significantly to the increase in the readout noise.

{
In order to solve the physical origin of readout noise increase, a more comprehensive analysis will be needed. We are now formulating the readout noise due to the $1/f$ noise and thermal noise generated in the MOSFET in the main amplifier in each pixel circuit. In the radiation environment, the $1/f$ noise increases due to the increase in the interface trap\cite{Nemirovsky2001}. Also, the increase in the sense-node capacitance\cite{hagino2020} must affect the propagation of these $1/f$ and thermal noise. These full noise analyses will be our future work.
}

\subsection{Dark Current}
\label{subsect:disc_dark}
The dark current increased by $89\pm 13\%$ at $10~\mathrm{krad}$ due to X-ray irradiation. We investigated its physical mechanism using TCAD device simulator HyDeLEOS, which is a part of the HyENEXSS~\cite{hyenexss}. In the simulation, we implemented the device structure as shown in Figure~\hyperlink{6C_crosssection}{\ref{6C_crosssection}} and calculated the dark current flowing in {a one-pixel region}. Detailed profiles for p-stops, sense nodes, BNWs, BPWs, and middle-Si layers were implemented based on the parameters provided by LAPIS Semiconductor Co. Ltd. In addition, it is generally known that the fixed positive charges are accumulated in the BOX layer during the wafer process. Therefore, we placed the fixed charge $Q_\mathrm{fix}$ of $2.0\times 10^{11}~\mathrm{cm^{-2}}$ uniformly between $1\textrm{--}3~\mathrm{nm}$ above the Si/SiO$_{2}$ interface in reference to a previous study~\cite{afanasev1998}.

We also implemented the radiation damage effects in the simulation. We reproduced the accumulation of positive charges by placing positive fixed charges $Q_\mathrm{BOX}$ in the BOX layer. We assumed that the concentration of $Q_\mathrm{BOX}$ increases in proportion to the dose based on the experimental results of a previous study~\cite{hara2019}. In addition, in order to consider the carrier generation through the interface traps, we used the surface recombination model as expressed by~\cite{smsze}
\begin{equation}
U_\mathrm{SUR}=\frac{n_\mathrm{i} ^{2}-pn}{(n+n_\mathrm{i})/S_\mathrm{p}+(p+n_\mathrm{i})/S_\mathrm{n}}~[\mathrm{cm^{-2}/s} ],
\label{sur}
\end{equation}
where $n_\mathrm{i}$ is the intrinsic carrier density, $p$ is the hole density, $n$ is the electron density, and $S$ is the surface recombination velocity. Subscripts p and n represent {hole and electron}, respectively. In this study, for the sake of simplification, we assume $S_\mathrm{n}=S_\mathrm{p}$. The surface recombination velocity is expressed as $S=\sigma v_\mathrm{th}N_\mathrm{it}$, where $\sigma$ is the carrier capture cross section, $v_\mathrm{th}$ is the thermal velocity, {and} $N_\mathrm{it}$ is the interface state density~\cite{smsze}. We can calculate the carrier generation rate through the interface traps by applying this model to the Si/SiO$_{2}$ interface. According to a previous study~\cite{shi2020}, $N_\mathrm{it}$ increases due to radiation damage, and thus the surface recombination velocity $S$ must increase after X-ray irradiation. In Eq. (\ref{sur}), $p$ and $n$ are calculated by device simulation and $n_\mathrm{i}$ is a constant. Therefore, since only $S_\mathrm{n,p}$ is an unknown parameter and depends on the dose, it is necessary to model $S_\mathrm{n,p}$ as a function of the dose based on the experimental results shown in Fig.~\hyperlink{Leak_Nro}{\ref{Leak_Nro}(a)}.

In order to model $S_\mathrm{n,p}$ as a function of the dose, we first need to reproduce the measured dark current in the simulation. Since the dark current is reproduced by the SUR (Eq. (\ref{sur})) and Shockley-Read-Hall (SRH) models~\cite{smsze} in the simulation, the unknown parameters, $S_\mathrm{n,p}$ in the SUR model and carrier lifetime $\tau_\mathrm{n,p}$ in the SRH model, are adjusted. In HyDeLEOS, it is possible to adjust the carrier lifetime $\tau_\mathrm{n,p}$ in the Si bulk by {the coefficients of the carrier lifetimes of electrons ($A_\mathrm{n}$) and holes ($A_\mathrm{p}$) as
\begin{equation}
\tau_\mathrm{n}=A_\mathrm{n}\times\tau_\mathrm{n0},~\tau_\mathrm{p}=A_\mathrm{p}\times\tau_\mathrm{p0}.
\label{tau}
\end{equation}
}
In this study, for the sake of simplification, we assumed $A_\mathrm{n}=A_\mathrm{p}$ as well as $S_\mathrm{n}=S_\mathrm{p}$. $\tau_\mathrm{n0}\simeq 12.9~\mathrm{\mu s}$ and $\tau_\mathrm{p0}\simeq 0.4~\mathrm{\mu s}$ are the fiducial values of the carrier lifetime defined as default parameters in HyDeLEOS for the sensor layer of XRPIX6C composed of p-type Si with a resistivity of $4~\mathrm{k\si{\ohm}~cm}$, which corresponds to the doping concentration of $3\times 10^{12}~\mathrm{cm^{-3}}$. 

\begin{figure}[tbp]
\centering
\begin{tabular}{c}
\includegraphics[width=\hsize]{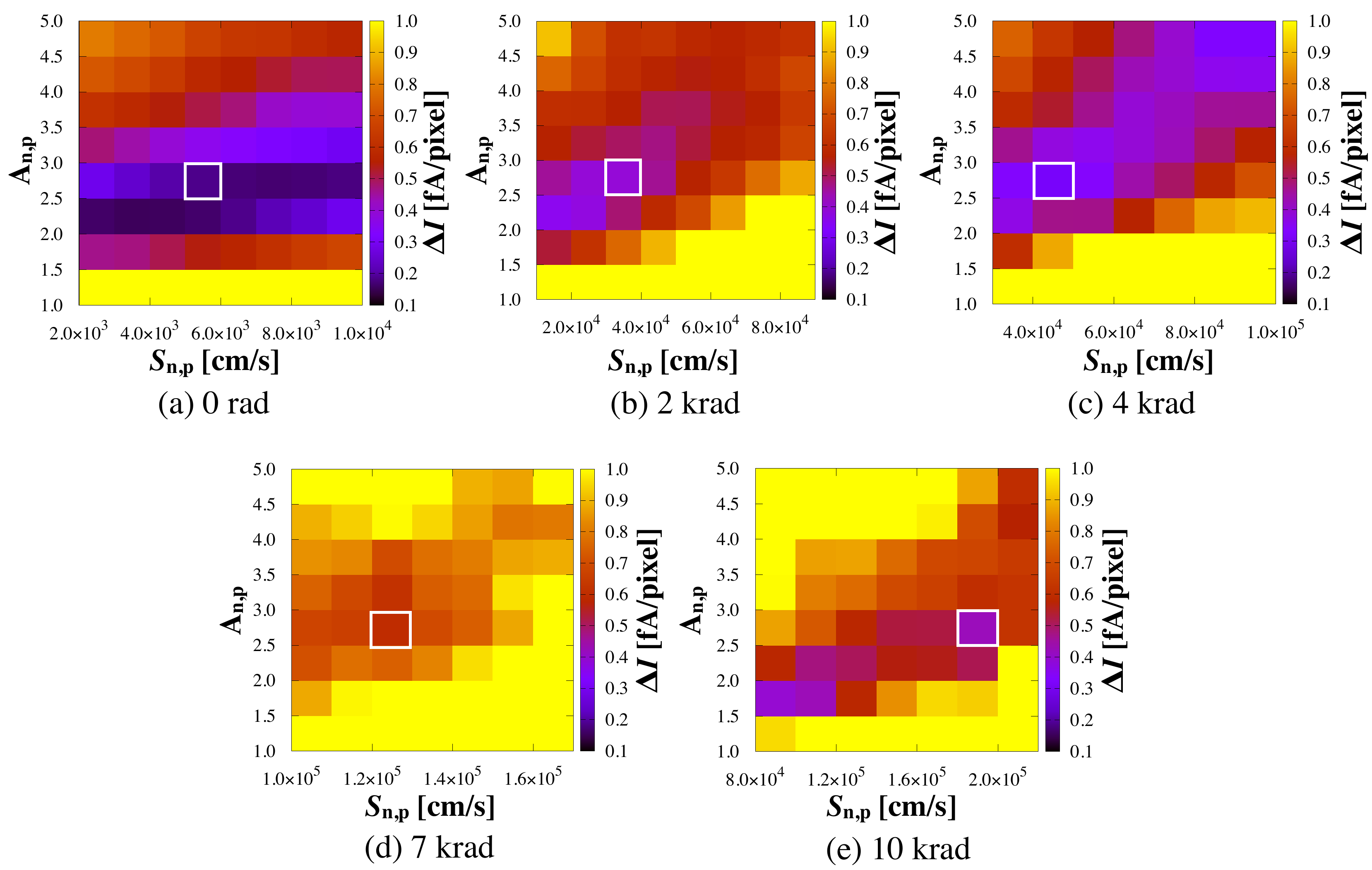}
\end{tabular}
\caption[The difference between the measured and simulated dark current $\Delta I$.]{The difference between the measured and simulated dark current $\Delta I$ as a function of the simulation parameters $S_\mathrm{n,p}$ and $A_\mathrm{n,p}$. The white squares indicate the best parameters which have a lowest $\Delta I$ for each dose.}
\label{param_tuning}
\end{figure}

In order to evaluate the reproducibility of the measured dark current by simulation, we calculated $\Delta I$, which is the average value of the {differences} between the measured and simulated dark {currents} at multiple {back-bias voltages} $V_\mathrm{BB}$. This takes into account the dark current generated from the sensor layer depleted by $V_\mathrm{BB}$. Then, we adjusted the parameters $S_\mathrm{n,p}$ and $A_\mathrm{n,p}$ to minimize $\Delta I$. As shown in Fig.~\hyperlink{param_tuning}{\ref{param_tuning}}, the optimal values of $A_\mathrm{n,p}$ do not change before and after the radiation damage because the lattice defects in Si bulk which shorten the carrier lifetime do not increase due to X-ray irradiation in principle. The carrier lifetime $\tau_\mathrm{n,p}$ remain at the same value ($\tau _\mathrm{n}\simeq 30~\mathrm{\mu s},~\tau _\mathrm{p}\simeq 1~\mathrm{\mu s}$) after $10~\mathrm{krad}$ irradiation. These values are reasonable compared to those measured by the microwave-detected photoconductance decay (MWPCD) method in a previous study~\cite{schroder1997}.  

\hypertarget{param_tuning}{}

\hypertarget{dose_optS}{}
\begin{figure}[tbp]
\centering
\begin{tabular}{c}
\includegraphics[width=0.8\hsize]{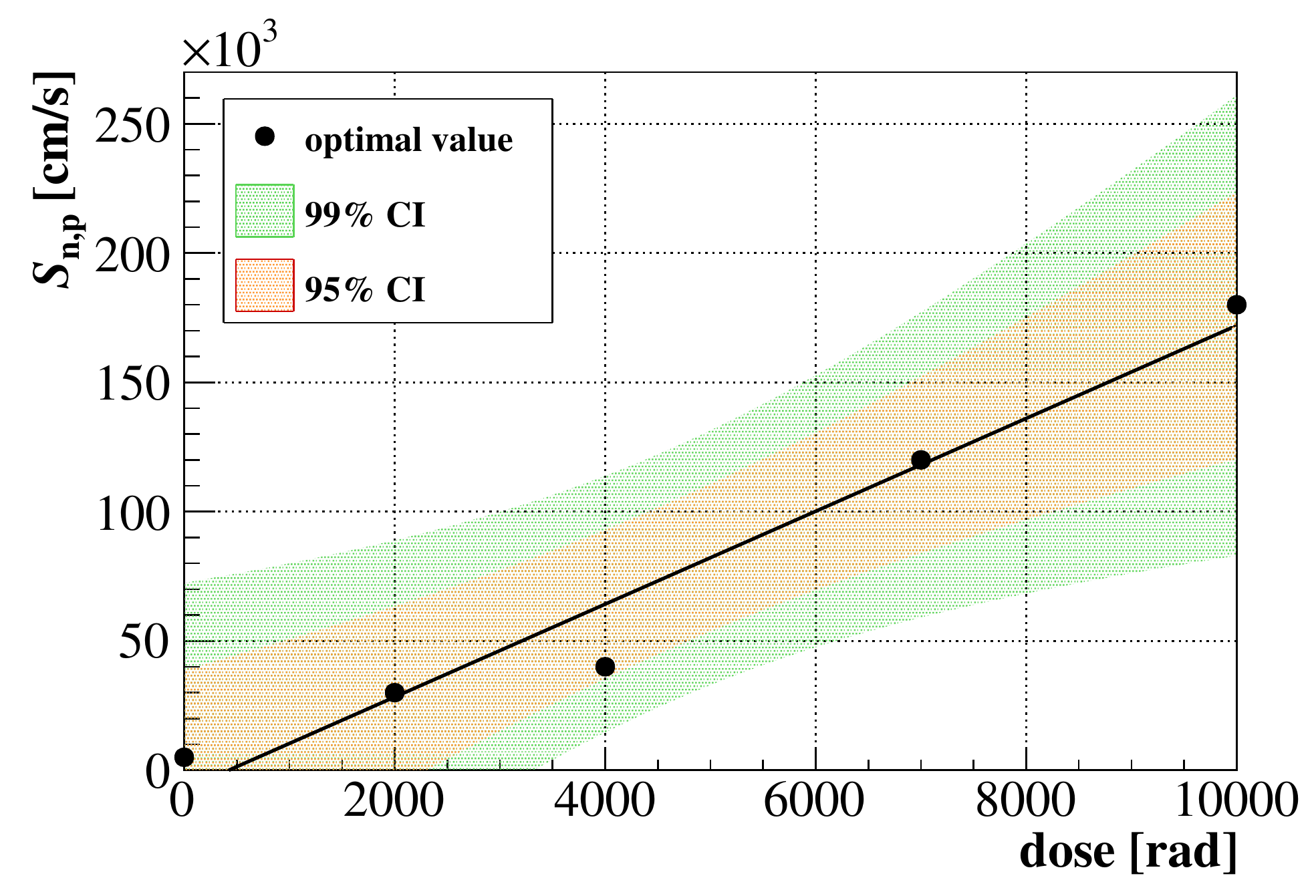}
\end{tabular}
\caption[The optimal values of $S_\mathrm{n,p}$ as a function of dose level.]{The optimal values of $S_\mathrm{n,p}$ as a function of dose level. The solid line indicates the best fit linear function. The shaded regions represent its confidence intervals for 95\% and 99\%.}
\label{dose_optS}
\end{figure}

On the other hand, the optimal value of $S_\mathrm{n,p}$ increased with increasing dose as shown in Fig.~\hyperlink{param_tuning}{\ref{param_tuning}}. Then, we modeled the dependence of $S_\mathrm{n,p}$ on the dose assuming a linear relationship between them. Figure~\hyperlink{dose_optS}{\ref{dose_optS}} shows the optimal values of $S_\mathrm{n,p}$ as a function of the dose, the best fit linear function and its confidence intervals for 95\% and 99\%. As a result of linear fitting, slope $18.0\pm2.1~\mathrm{cm\cdot s^{-1} \cdot rad^{-1}}$ and intercept $(-7.6\pm1.2)\times10^{3}~\mathrm{cm/s}$ were obtained as the best-fit parameters. According to the obtained linear model, the surface recombination velocity is $S_\mathrm{n,p}\simeq 1.7\times 10^{5}~\mathrm{cm/s}$ after 10 krad irradiation. This value of the surface recombination velocity is consistent with a previous study~\cite{tonigan2021} for an irradiation of 10 krad.

\hypertarget{dose_leak_sim-expt}{}
\begin{figure}[tbp]
\centering
\begin{tabular}{c}
\includegraphics[width=0.8\hsize]{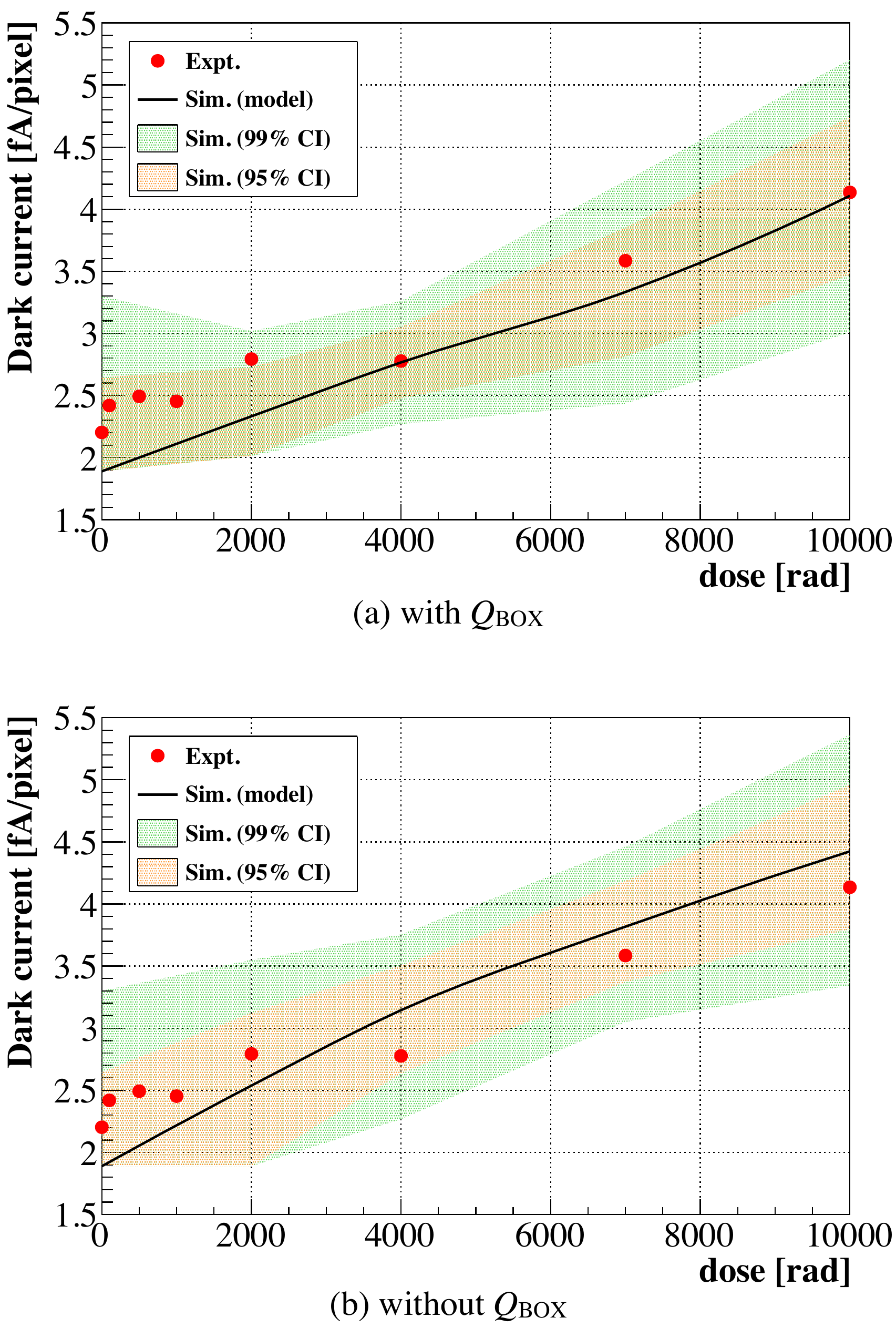}
\end{tabular}
\caption[Comparison of the experimental dark current degradation with the simulation.]{Comparison of the experimental dark current degradation with the simulation in the case (a) with $Q_\mathrm{BOX}$ and (b) without $Q_\mathrm{BOX}$. The solid lines show the simulated dark current using the assumed linear model. The shaded regions show the simulated dark current using $S_\mathrm{n,p}$ at the upper and lower limits of the confidence intervals for each dose in Fig.~\hyperlink{dose_optS}{\ref{dose_optS}}. }
  \label{dose_leak_sim-expt}
\end{figure}

Using the linear model of $S_\mathrm{n,p}$ against the dose shown in Fig.~\hyperlink{dose_optS}{\ref{dose_optS}}, we compare the dose dependence of the measured dark current with the simulated dark current. Figure~\hyperlink{dose_leak_sim-expt}{\ref{dose_leak_sim-expt}(a)} shows the comparison between the measured dark current and simulated dark current using the assumed linear model. The shaded regions show the simulation results corresponding to the confidence intervals for each dose in Fig.~\hyperlink{dose_optS}{\ref{dose_optS}}. However, since $S_\mathrm{n,p}$ never takes negative values, $S_\mathrm{n,p}$ at 0 rad of the linear model and the lower limits of the confidence intervals at 0 rad and 2 krad are set to $0~\mathrm{cm/s}$. As shown in Fig.~\hyperlink{dose_leak_sim-expt}{\ref{dose_leak_sim-expt}(a)}, TCAD simulation successfully reproduced the experimental result by taking into account two radiation damage effects, i.e., accumulation of BOX charges and increase of interface traps. 

In order to understand how the accumulated positive charge $Q_\mathrm{BOX}$ and the interface traps contribute to the dark current, we discriminate these effects. In Fig.~\hyperlink{dose_leak_sim-expt}{\ref{dose_leak_sim-expt}(b)}, $Q_\mathrm{BOX}$ is not added at all for any doses. In both Fig.~\hyperlink{dose_leak_sim-expt}{\ref{dose_leak_sim-expt}(a)} and~\hyperlink{dose_leak_sim-expt}{\ref{dose_leak_sim-expt}(b)}, the parameter $S_\mathrm{n,p}$ increases approximately in proportion to the dose as shown in Fig.~\hyperlink{dose_optS}{\ref{dose_optS}}, and the simulated dark current tends to increase as the dose increases. Therefore, the increase in the interface traps contributes to the dark current increase.

\hypertarget{elec_qbox}{}
\begin{figure}[tbp]
\centering
\begin{tabular}{c}
\includegraphics[width=0.8\hsize]{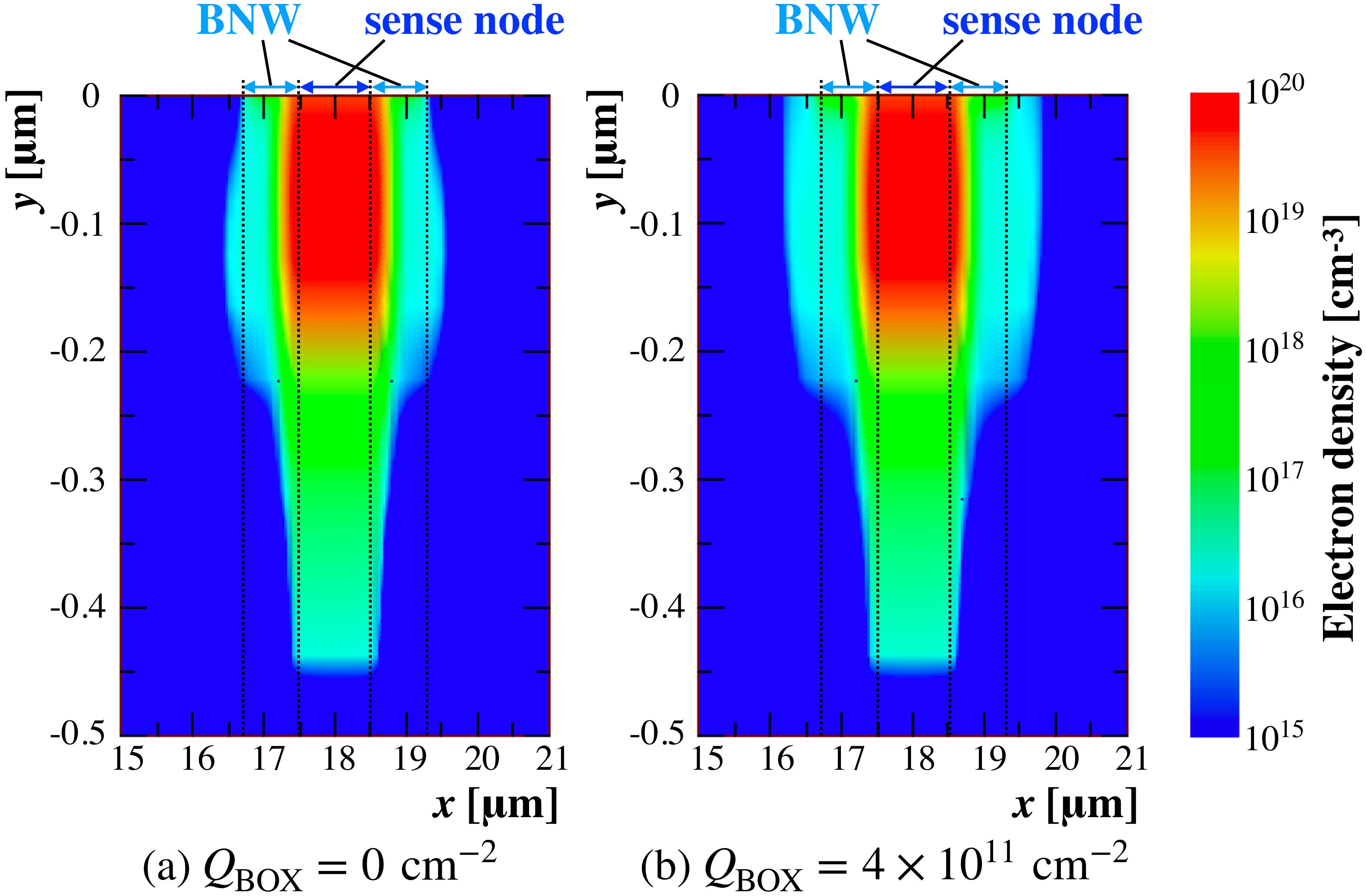}
\end{tabular}
\caption[The electron density distribution near the Si/SiO$_{2}$ interface ($y = 0$).]{The electron density distribution near the Si/SiO$_{2}$ interface ($y = 0$). The left panel and right panel are the case with and without $Q_\mathrm{BOX}$, respectively.}
\label{elec_qbox}
\end{figure}

Comparing Fig.~\hyperlink{dose_leak_sim-expt}{\ref{dose_leak_sim-expt}(a)} and~\hyperlink{dose_leak_sim-expt}{\ref{dose_leak_sim-expt}(b)}, the simulated dark current in Fig.~\hyperlink{dose_leak_sim-expt}{\ref{dose_leak_sim-expt}(a)}, which contains $Q_\mathrm{BOX}$ with different concentrations at each dose, is slightly lower than that in Fig.~\hyperlink{dose_leak_sim-expt}{\ref{dose_leak_sim-expt}(b)}, which contains no $Q_\mathrm{BOX}$ at all. In order to reveal {the physical mechanism of this difference}, we focus on the electron density distribution near the Si/SiO$_{2}$ interface. Fig.~\hyperlink{elec_qbox}{\ref{elec_qbox}(a)} and~\hyperlink{elec_qbox}{\ref{elec_qbox}(b)} show the electron density map around the BNW without $Q_\mathrm{BOX}$ and with $Q_\mathrm{BOX}$, respectively. In the case without $Q_\mathrm{BOX}$ of Fig.~\hyperlink{elec_qbox}{\ref{elec_qbox}(a)}, both ends of the BNW are depleted and the Si/SiO$_{2}$ interface is bare. On the other hand, in the case with $Q_\mathrm{BOX}$ of Fig.~\hyperlink{elec_qbox}{\ref{elec_qbox}(b)}, the region of high electron density extends horizontally because the $Q_\mathrm{BOX}$ attracts electrons near the Si/SiO$_{2}$ interface. These electrons fill the Si/SiO$_{2}$ interface, making it difficult for carriers to be generated, resulting in lower dark current. Therefore, $Q_\mathrm{BOX}$ does not increase the dark current, but slightly decrease it.

\section{Conclusion}
\label{sect:conc}
We performed an irradiation experiment on D-SOI XRPIX using $\sim 10~\mathrm{keV}$ X-rays with a total dose of $10~\mathrm{krad}$ and investigated the physical mechanism of the degradation of detector performance.
{As the results, we found that the energy resolution at 5.9 keV X-ray degraded by 17.8 ± 2.8\%, and the dark current increased by 89 ± 13\%.}
Especially regarding the dark current, we found that the increase in the interface trap density predominantly contributes to the increase in it. Moreover, {the} accumulated positive charge in the BOX layer does not increase the dark current. {Thus, in the case of XRPIX and possibly the other SOI pixel sensors as well, it is important to reduce the dark current due to the interface traps in order to suppress the increase in dark current under the radiation environment.}




\section* {Acknowledgments}
We acknowledge the relevant advice and manufacture of the XRPIXs by the personnel of LAPIS Semiconductor Co., Ltd. This study was supported by MEXT/JSPS KAKENHI Grant-in-Aid for Scientific Research on Innovative Areas 25109002 (Y.A.) and 25109004 (T.G.T., T.T., K.M., A.T., and T.K.), Grant-in-Aid for Scientific Research (B) 25287042 (T.K.), Grant-in-Aid for Young Scientists (B) 15K17648 (A.T.), Grant-in-Aid for Challenging Exploratory Research 26610047 (T.G.T.), and Grant-in-Aid for Early Career Scientists 19K14742 (A.T.). This study was also supported by the VLSI Design and Education Center (VDEC), Japan, and the University of Tokyo, Japan in collaboration with Cadence Design Systems, Inc., USA; Mentor Graphics, Inc., USA; and Synopsys, Inc, USA. We also thank Kazuya Matsuzawa, Yutaka Akiyama, and Nobutoshi Aoki (Kioxia corp.) for the fruitful discussions on the device simulation.



\bibliography{report}   
\bibliographystyle{spiejour}   


\vspace{2ex}\noindent\textbf{Masatoshi~Kitajima} is a master's student at Tokyo University of Science in Japan. He received his BS degree in physics from Tokyo University of Science in 2020. His current research interest includes developments of X-ray SOI pixel detectors for the X-ray astronomical satellite.


\vspace{1ex}
\noindent

\listoffigures
\listoftables

\end{spacing}
\end{document}